# High-Q Fano resonance in all-dielectric metasurfaces for molecular fingerprint detection

**S. HADI BADRI[1], M. M. GILARLUE[1], SANAM SAEIDNAHAEI[2,*], JONG SU KIM[2,*]**

[1] *Department of Electrical Engineering, Sarab Branch, Islamic Azad University, Sarab, Iran.*
[2] *Department of Physics, Yeungnam University, Gyeongsan, 38541, Republic of Korea.*
[*] *sanam.nahaei@yu.ac.kr (S. SaeidNahaei) and jongsukim@ynu.ac.kr (J. S. Kim)*

**Abstract:** We present and numerically investigate a high-quality factor (high-Q) meta-atom with Fano resonance. Numerical simulations indicate that the designed meta-atom has a single sharp Fano resonance in the 1350-1750 1/cm range. Moreover, the frequency of the single resonance can be tuned in this frequency range by scaling the meta-atom. We exploit these properties to design a pixelated metasurface for spectrometer-less molecular fingerprint retrieval. The proposed meta-atom with an average quality factor of 2000 makes it possible to decrease the scaling step of metapixels without introducing any resonance overlap between the metapixels leading to higher precision in label-free and non-destructive identification of the molecular fingerprints.

## 1. Introduction

Metamaterials provide unique opportunities to manipulate electromagnetic fields [1] and enable us to implement unique photonic devices [2-6]. Metasurfaces, the two-dimensional equivalents of metamaterials, have a remarkable ability for controlling the flow of light [7, 8]. Fano-resonant metasurfaces are of great interest considering their narrow spectral linewidth. Different applications such as sensing [9-11], lasing [12, 13], switching [14, 15], wavelength demultiplexing [16, 17], and nonlinear applications [18] have been introduced based on Fano-resonant metasurfaces. The Fano resonance with a distinctly asymmetric shape originates from the interference of interacting resonances [19]. The Fano resonances have been observed in various structures such as photonic crystal slab [20], subwavelength hole arrays [21], plasmonic nanocavities [22], and nanoparticle arrays [23].

Spectroscopy techniques have been used to identify molecular species based on their characteristic absorption fingerprints due to their vibrational modes and absorption band of their chemical bonds. spectroscopic equipment such as Fourier-transform infrared (FTIR) spectrometers or tunable laser-based light sources is bulky and expensive. Moreover, bulk absorption spectroscopy has its own limitations imposed by the exponential decrease of the vibrational signal with the analyte thickness [24]. The thickness of trace-amount substances is typically much smaller than the mid-infrared (mid-IR) wavelengths leading to considerably weak light-matter interaction and, consequently, lower precision in the identification of the molecular fingerprints [25]. Spectrometer-less biosensors based on high-Q metasurfaces can overcome the mentioned limitations for detecting complex molecular species. A pixelated dielectric metasurface, where each metapixel is engineered to have a single high-Q resonance in the desired frequency range, has been proposed [26]. Each metapixel consists of a zigzag array of elliptical resonators where the resonance is attributed to the quasi-bound state in the continuum (quasi-BIC) formed by breaking the in-plane inversion symmetry of the unit cell [27]. By linearly scaling the unit cell and the resonators, each metapixel can be tuned to a distinct resonance frequency. This one-to-one mapping of each resonance to a specific pixel in the metasurface enables us to correlate the modulation of the individual metapixel resonance to absorption bands of the analyte molecules. The coupling between vibrational modes of the molecules and the enhanced field on the surface of the metapixel modulates the reflectance spectrum which is observed as attenuation and broadening of the metapixel resonance. Each metapixel's reflectance is recorded before and after coating the metasurface with the analyte through image-based readouts. Finally, the changes in the reflectance of the metapixels are translated to the absorption signature of the analyte. Pixelated metasurfaces have also been utilized to implement other functionalities such as beam deflection [28], polarization detection [29], polarization cameras [30], displays [31, 32], energy harvesting [33], and imaging [34, 35]. Another absorption fingerprint retrieval technique has been proposed based on the incidence angle-scanning of a single metasurface. Controlling the incidence angle of the source enables us to get a large number of resonances from a metasurface reducing the footprint of the sensor. By comparing the reflectance spectrum of the metasurface at the presence and the absence of the analyte, the absorption signature of the analyte is calculated [36].

In this paper, we present a high-Q meta-atom and investigate its application in pixelated metasurfaces for molecular fingerprint retrieval without requiring spectrometry, frequency scanning, or moving mechanical components. The designed meta-atom is compatible with complementary metal-oxide-semiconductor (CMOS) technology and is composed of two dielectric rectangles where a small rectangular shape is removed from the inner sides of the resonators. The average quality factor of the proposed metapixels is about 2000 while the average quality factor of the zig-zag arrays of elliptical metapixels presented in [26] is about 200. The exposure of the pixelated metasurface to the analyte results in the broadening and attenuation of the resonances, therefore, metapixels with higher quality factors provide higher accuracy in image-based readouts. The scaling steps in designing the metapixels with low quality factor is limited due to the overlap of the resonances of the metapixels with small scaling steps. However, our designed meta-atom with much narrower linewidth can be scaled by considerably smaller steps leading to higher precision in the readouts.

## 2. Meta-atom for pixelated metasurface

A suitable meta-atom for a pixelated metasurface should have high quality factor. Besides, the reflectance spectrum of the meta-atom should be spectrally clean, i.e., without additional resonances and should have a single resonance in the desired frequency range. The proposed meta-atom with the mentioned design objectives is shown in Fig. 1. The designed meta-atom is composed of two hydrogenated amorphous silicon (a-Si:H) rectangles with a length and width of $l$=1.9 μm and $w$=1.4 μm, respectively. A rectangular indentation with a size of $l_r=\alpha_1\times l$ and $w_r= \alpha_2\times w$, where $\alpha_1$=0.1 and $\alpha_2$=0.6, is created in the inner sides of the meta-atoms. The a-Si:H patches with a thickness of $h$=0.7 μm are placed on an $MgF_2$ substrate with a thickness of $h_{sub}$=0.7 μm. The unit cell's periodicity is defined as $P_x$=2.2×$l$=4.18 μm and $P_y$=2.2×$w$=3.08 μm. The gap between the rectangles is $g=P_x/2-l$. And $S$ is the scaling factor which is used to scale the meta-atom which in turn shifts the resonance frequency of the metasurface. The numerical simulations of the designed metasurface were performed using the finite-element method (FEM) available in commercial CST STUDIO SUITE software. The average refractive index of the a-Si:H is 3.21 while the refractive index of $MgF_2$ is considered to be 1.31. The extinction coefficients for both materials were taken as $k$=0 [26]. In our calculations, we consider a homogenous background with a permittivity of unity. A plane wave polarized along the $y$-axis is incident normally to the metasurface. Periodic boundary conditions are applied along the $x$- and $y$-axes while in the $z$-axis the perfectly matched layer (PML) boundary conditions are applied.

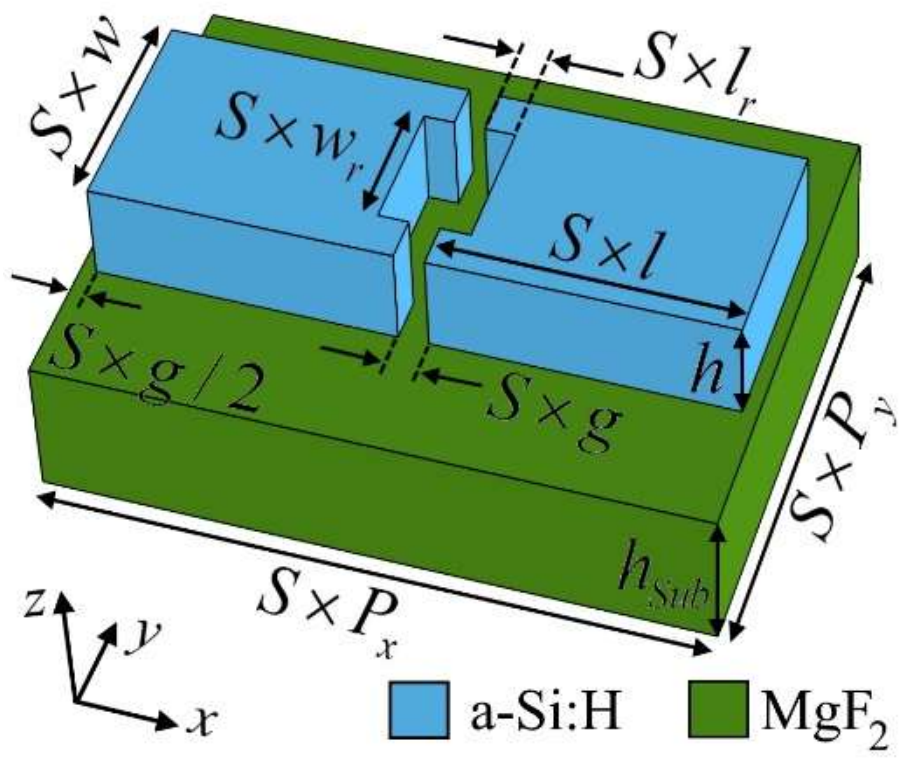

Fig. 1. Schematic of the designed unit cell of the pixelated metasurface where *S* is the scaling factor.

The reflectance spectrum of the proposed meta-atom with the scaling factor of $S$=1 is shown in Fig. 2. The reflectance of the structure is calculated as $R=|S_{11}|^2$. We should mention that higher quality factor can be achieved by optimizing the size of the defect in the proposed meta-atom. However, in the reflectance spectrum of such meta-atoms or their scaled version, multiple resonances appear in the desired frequency range making such meta-atoms unsuitable for pixelated metasurfaces. To show that the observed asymmetric line-shape resonance is a Fano resonance, we compare the calculated resonance with the Fano formula. As shown in Fig. 2, the reflectance spectrum of the metasurface is fitted by using the typical Fano formula [37]

$$R_{Fano}(\omega)=\left|a_1+ja_2+\frac{b}{\omega-\omega_0+j\gamma}\right|^2 \tag{1}$$

where $a_1$, $a_2$, and $b$ are constant real numbers while $\omega_0$ is the resonant frequency. The overall damping rate of the resonance, $\gamma$, is proportional to the line-width of the resonance. The quality factor of the meta-atom with the scaling factor of $S$=1 is $Q=\omega_0/2\gamma$ which is as high as 2120.

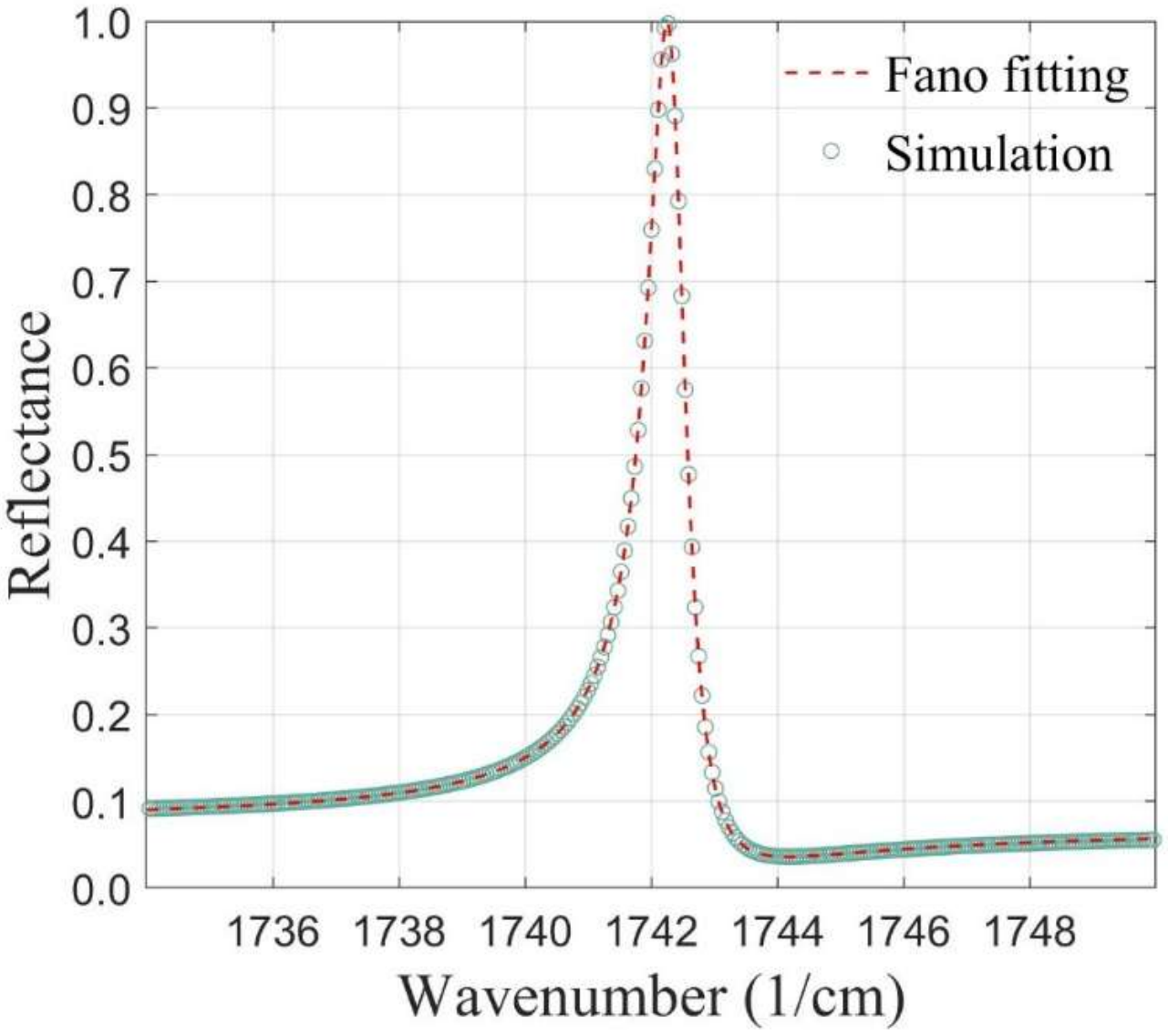

Fig. 2. The reflectance spectrum of a metasurface with a scaling factor of $S$=1. The normally incident electric field is polarized along the $y$-axis.

We investigate the physical mechanism of the High-Q Fano resonance of the metasurface by performing the Cartesian multipole decomposition [38-42]. The induced current density, $\boldsymbol{J}(\boldsymbol{r})$, in the meta-atom can be calculated by using the electric field distribution, $\boldsymbol{E}(\boldsymbol{r})$, and the refractive index distribution, $n$ [43].

$$\boldsymbol{J}(\boldsymbol{r}) = -i\omega\varepsilon_0(n^2-1)\boldsymbol{E}(\boldsymbol{r}) \tag{2}$$

where $\boldsymbol{r}$ is a position vector from the origin to point $(x, y, z)$, $\omega$ is the angular frequency, and $\varepsilon_0$ is the permittivity of the free space. Here, we consider the dominant terms in the multipole decomposition. The electric dipole ($\boldsymbol{p}$), magnetic dipole ($\boldsymbol{m}$), electric quadrupole ($Q^e$), and magnetic quadrupole ($Q^e$) are defined as [38]

$$\boldsymbol{P} = \frac{1}{i\omega}\int \boldsymbol{j} d^3r \tag{3a}$$

$$\boldsymbol{M} = \frac{1}{2c}\int (\boldsymbol{r}\times\boldsymbol{j}) d^3r \tag{3b}$$

$$\boldsymbol{Q}^e_{\alpha\beta} = \frac{1}{2i\omega}\int \left[ r_\alpha j_\beta + r_\beta j_\alpha - \frac{2}{3}(\boldsymbol{r}\cdot\boldsymbol{j})\delta_{\alpha,\beta} \right] d^3r \tag{3c}$$

$$\boldsymbol{Q}^m_{\alpha\beta} = \frac{1}{2c}\int \left[ (\boldsymbol{r}\times\boldsymbol{j})_\alpha r_\beta + (\boldsymbol{r}\times\boldsymbol{j})_\beta r_\alpha \right] d^3r \tag{3d}$$

where c is the speed of light and $\alpha, \beta = x, y, z$. The effect of toroidal dipole and other higher-order modes is negligible on the scattered intensity. Therefore, the total scattering cross-section is the sum of these multipole moments [44, 45]

$$\begin{aligned}\sigma_{sca}^{total} &= \sigma_{sca}^{p} + \sigma_{sca}^{m} + \sigma_{sca}^{Q^e} + \sigma_{sca}^{Q^m} + \ldots \\ &= \frac{k^4}{6\pi\varepsilon_0^2\left|\boldsymbol{E}_{inc}\right|^2}\left[\sum_\alpha\left(\left|p_\alpha\right|^2 + \frac{\left|m_\alpha\right|^2}{c}\right) + \frac{1}{120}\sum_{\alpha\beta}\left(\left|kQ^e_{\alpha\beta}\right|^2 + \left|\frac{kQ^m_{\alpha\beta}}{c}\right|^2\right) + \ldots\right.\end{aligned} \tag{4}$$

where $\boldsymbol{E}_{inc}$ is the amplitude of the incident plane wave and $k$ is the wavenumber. We employed Lumerical FDTD solutions software to export the refractive index and electric field distribution to MATLAB where the induced current density is calculated by Eq. 2. Next, we calculate multipole moments by Eq. 3 and with a tool introduced in [43]. In the Lumerical, the periodic boundary conditions are applied along the $x$- and $y$-axes while the perfectly matched layer (PML) is applied along the $z$-axis. The mesh resolution is set to 3 under the auto-nonuniform type defined in the Lumerical. The calculated multipole moments are shown in Fig. 3. As can be seen in this figure, the strongest multipoles are magnetic dipole and electric quadrupole. The resonance frequency is located at 2027.3 1/cm for the scaling factor of $S$=1. The difference in the resonance frequency stems from the difference in the calculation methods of CST and Lumerical which are FEM and FDTD, respectively. Moreover, the simulation settings such as mesh and PML are different in these simulations. The electric and magnetic fields in the unit cell with a scaling factor of $S$=1 at the resonance frequency are displayed in Fig. 4. The magnetic field direction in the $x$-$z$ plane is shown in Fig. 4(b) corresponding to the dominant magnetic dipole. The $E_x$ component in the $x$-$y$ plane and at the interface of the resonators and the substrate is shown in Fig. 4(c) corresponding to the relatively weaker electric quadrupole. The minus and plus signs indicate the negative and positive charge distributions. To further explore the resonance mechanism of the metasurface, we also illustrate the magnetic and electric fields at the frequencies below and above resonance wavenumber, i.e., about 1737 and 1745 1/cm in the inset of Fig. 5. The magnetic dipole and the electric quadrupole at these frequencies are antiphase. The interference between these antiphase modes results in the appearance of the sharp Fano resonance with asymmetric spectral lineshape at 1742.2 1/cm [10, 46-48].

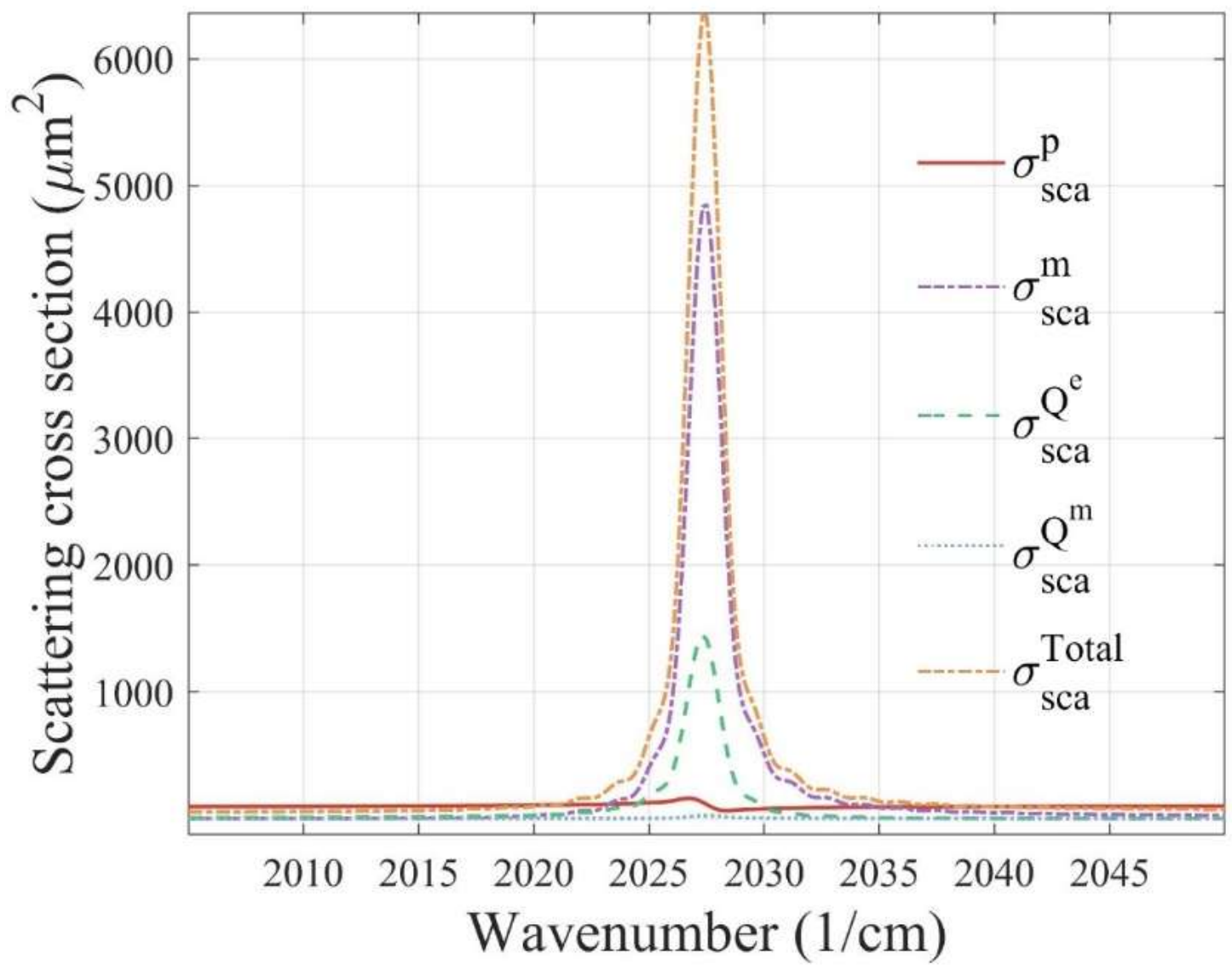

Fig. 3. The multipole decomposition method determining the contribution of each multipole moment.

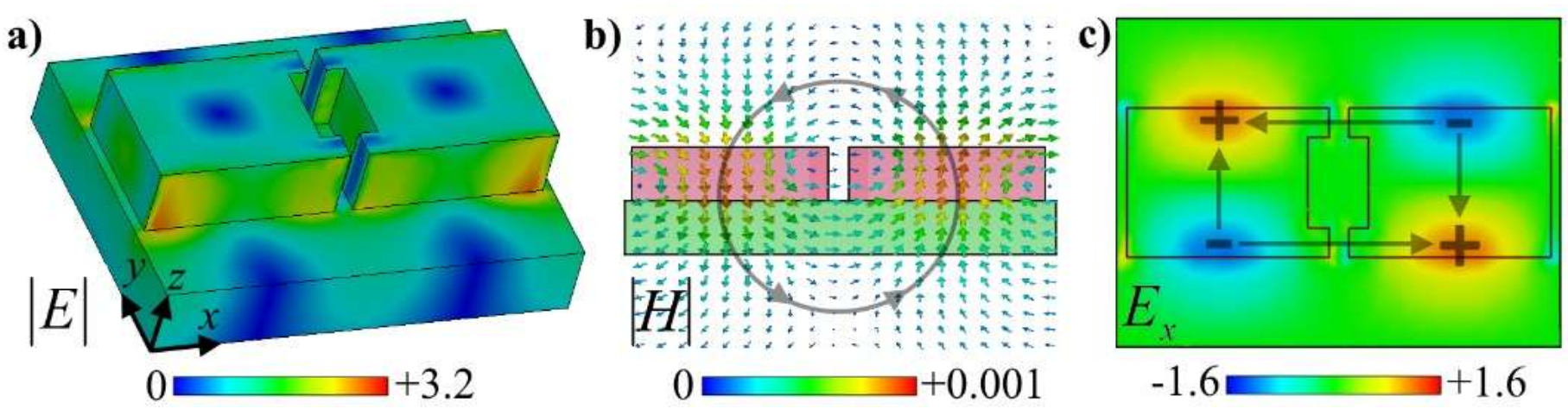

Fig. 4. Distribution of a) the electric field intensity, b) the magnetic field direction in the *x-z* plane showing the dominant magnetic dipole, and c) the electric field's $E_x$ component in the *x-y* plane showing the electric quadrupole at the resonance frequency.

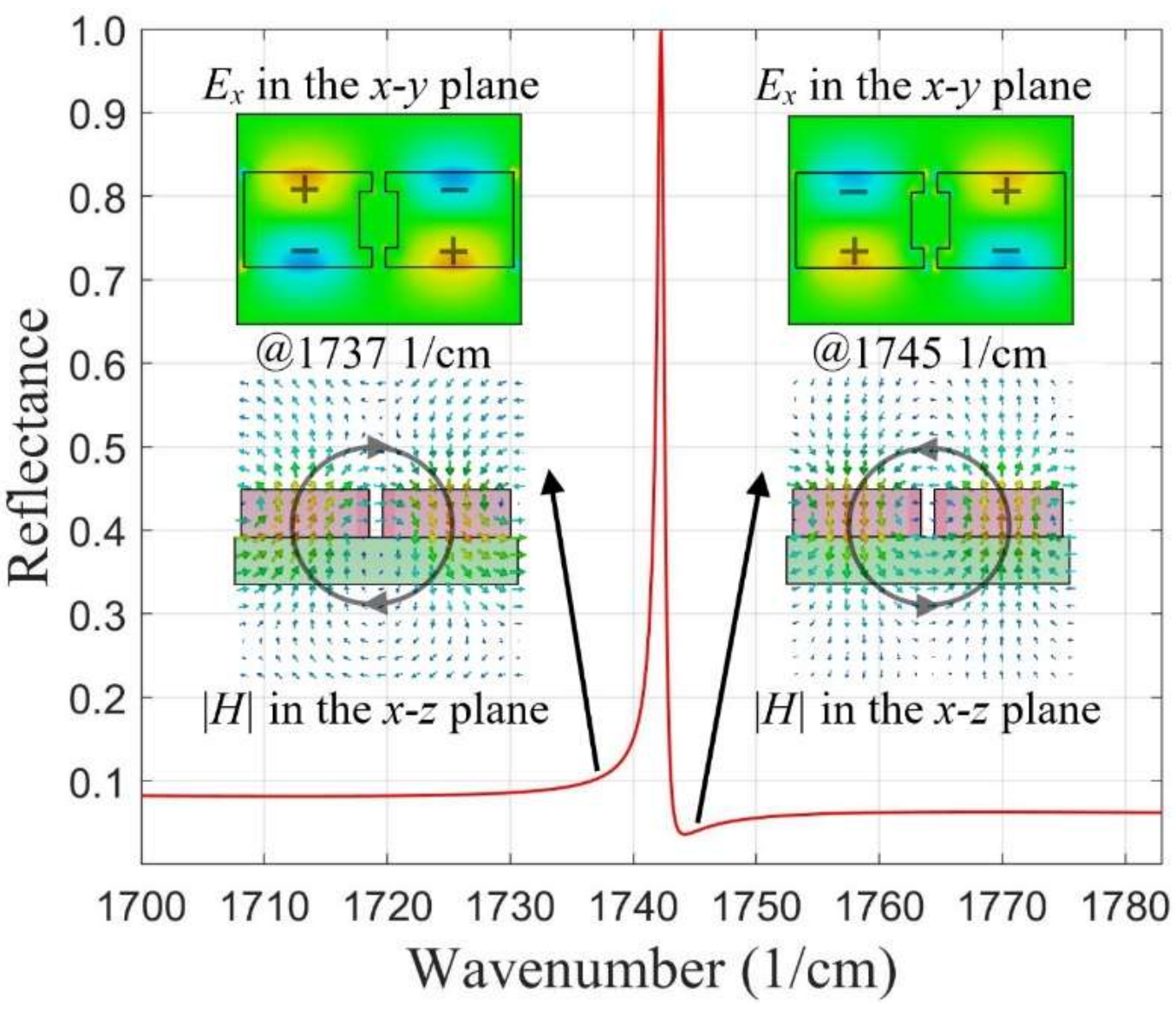

Fig. 5. The reflectance spectrum and the antiphase modes at the frequencies below and above resonance wavenumber, i.e., about 1737 and 1745 1/cm. The scaling factor is $S$=1.

## 3. Effect of geometrical parameters

In this section, we briefly study the effect of the geometrical parameters introduced in Fig. 1 on the Fano resonance of the proposed meta-atom. As discussed in the previous section, our goal is to design a high-quality factor meta-atom with a single resonance in the desired frequency range of 1300-1850 1/cm. Moreover, the scaled meta-atom should also have a single resonance in this frequency range. This is essential in designing a pixelated metasurface. The rectangular indentation in the inner sides of the meta-atom plays an important role in the resonance. The parameters $\alpha_1$ and $\alpha_2$ determine the size of this rectangular indentation, i.e., $l_r=\alpha_1\times l$ and $w_r=\alpha_2\times w$. First, we fix the other geometrical parameters and only change $\alpha_1$ from 0.05 to 0.2 for different scaling factors. As can be seen in Fig. 6, decreasing the length of the rectangular indentation leads to a higher quality factor. For instance, the quality factors of about 6000, 1890, 800, and 370, for the meta-atom with a scaling factor of $S$=1.34, are achieved corresponding to $\alpha_1$=0.05, 0.10, 0.15, and 0.20, respectively. It should also be noted that the resonance shifts to lower frequencies as the length of the rectangular indentation decreases. While $\alpha_1$=0.05 offers a considerably higher quality factor, there is no resonance corresponding to meta-atoms with scaling factors of $S$=1.00 and $S$=1.17. Therefore, we chose $\alpha_1$=0.10. We also tune the width of the rectangular indentation by changing $\alpha_2$ from 0.2 to 0.8 as shown in Fig. 7. In this case, the meta-atoms with $\alpha_2$=0.6 only exhibit a resonance in the desired frequency range for different scaling factors. While for other values of $\alpha_2$, the meta-atoms do not have resonance in the desired frequency range for different scaling factors. Therefore, meta-atoms with other values of $\alpha_2$ cannot be used as a pixelated metasurface. The effect of the substrate's thickness on the resonance of the structure is examined in Fig. 8. Decreasing the thickness of the substrate from $h_{sub}$=1.1 to 0.5 μm blueshifts the resonance frequency from 1733.4 to 1749.1 1/cm for a meta-atom with a scaling factor of unity. Moreover, as the thickness decreases the line-shape becomes more asymmetric. The reflection spectrum of a metasurface without substrate is also shown in Fig. 8.

We should also point out that our results are based on numerical simulations while [26] offers experimental results. Nevertheless, the quality factor of the fabricated metasurface could degrade from the values calculated by numerical simulations due to fabrication imperfections and surface roughness. We numerically investigate the rounding of the meta-atom's edges with a scaling factor of unity. Rounding with a fillet radius of 0.1 μm results in the blue-shift of 1.0 1/cm while the change in the quality factor of the metasurface is infinitesimal.

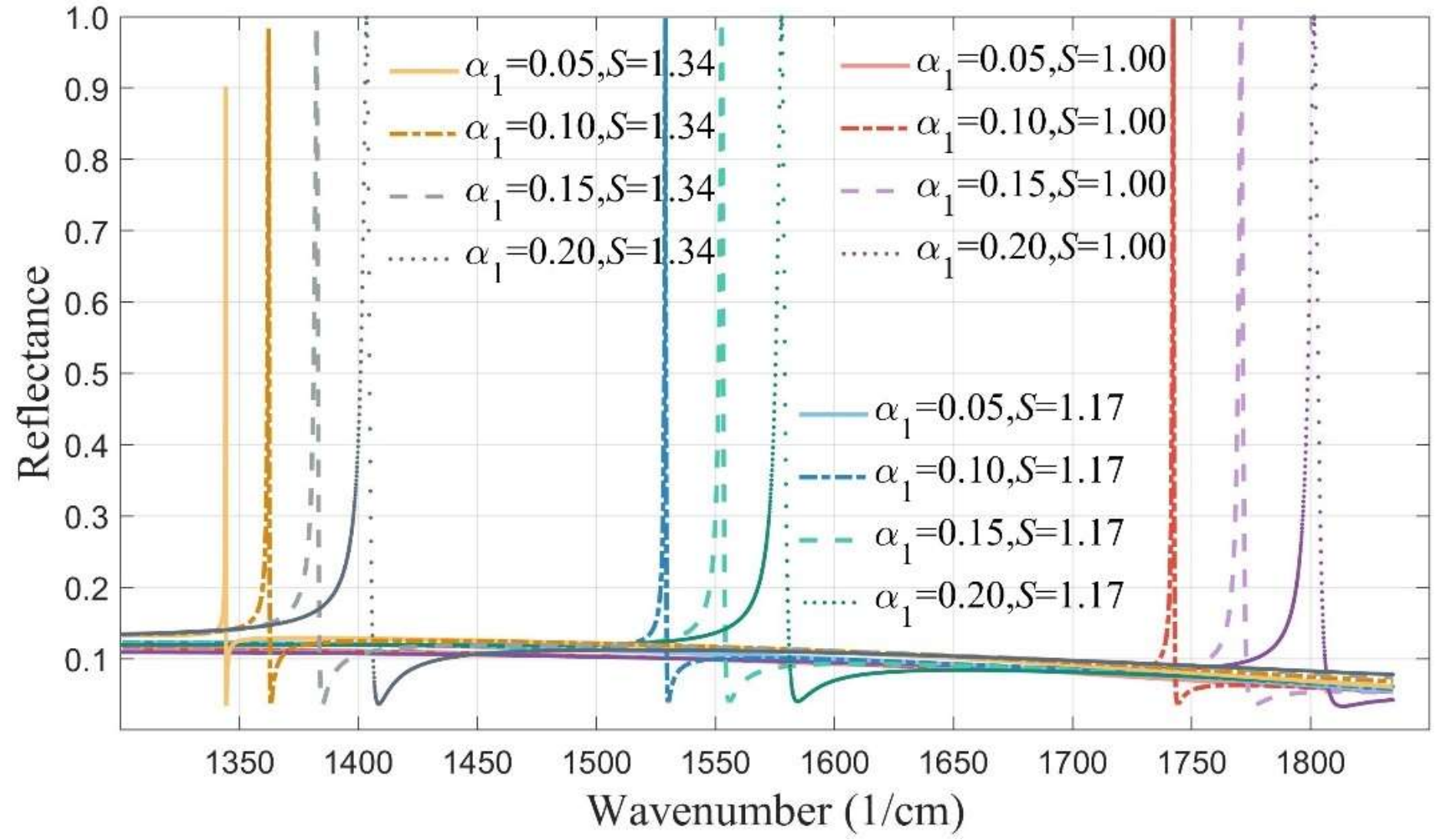

Fig. 6. The effect of rectangular indentation's length on the resonance of the scaled meta-atom. In some cases, there is no resonance in the desired frequency range.

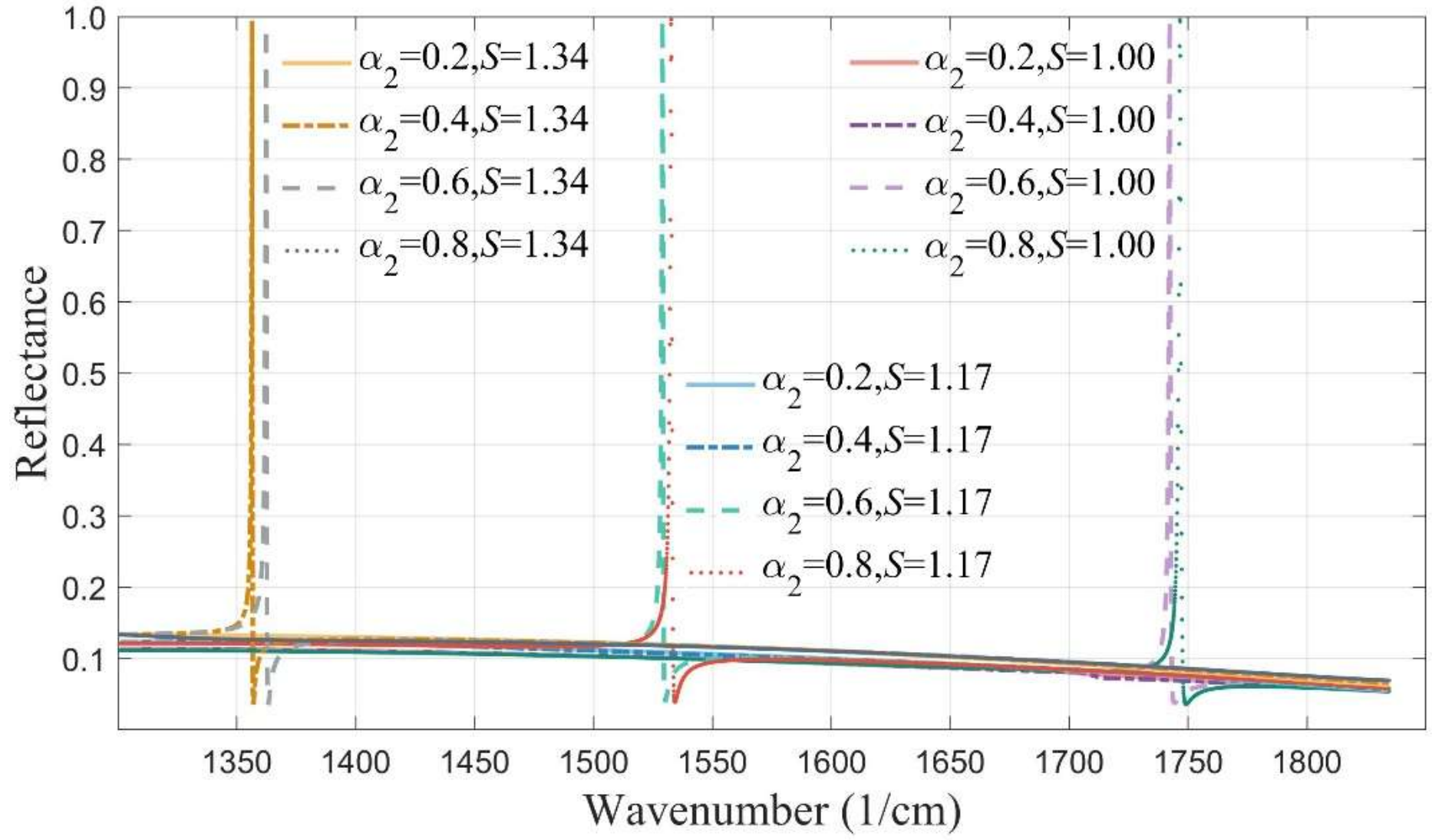

Fig. 7. The effect of the rectangular indentation's width on the resonance of the scaled meta-atom. In some cases, there is no resonance in the desired frequency range.

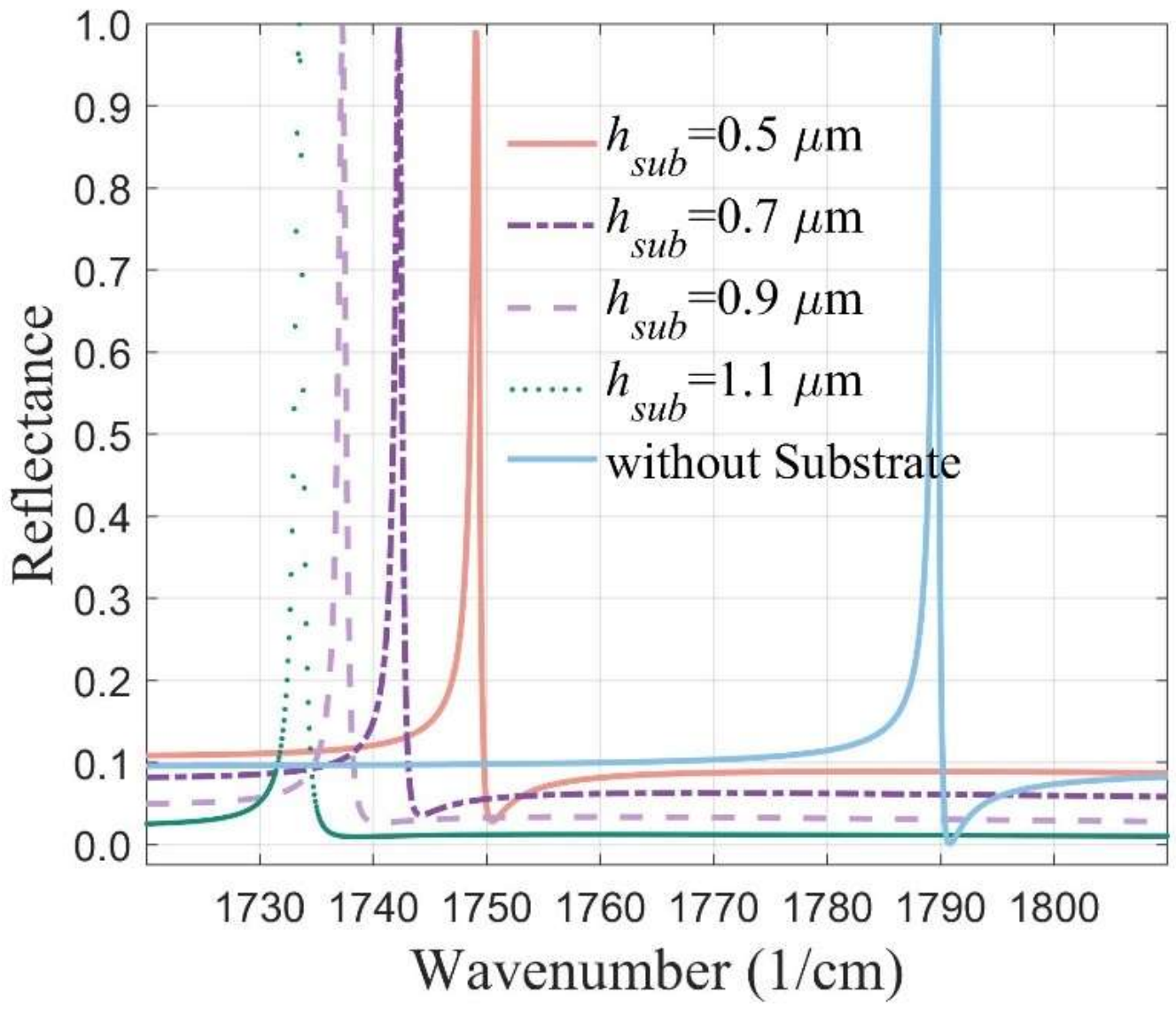

Fig. 8. The effect of the substrate's thickness on the resonance of the meta-atom with a scaling factor of $S$=1.

## 4. Pixelated metasurface for molecular fingerprint retrieval

The reference reflectance spectra of the metasurface for metapixels without the presence of the analyte are illustrated in Fig. 9. For illustration purposes, only 12 metapixles are shown in this figure. In each metapixel, the meta-atom and the unit cell are scaled by a factor of $S$. Therefore, each metapixel has a single resonance at a unique frequency compared to other metapixels. By carefully choosing a scaling step and the number of metapixels, we can obtain sharp resonances in a wide spectral range of 1350 to 1750 1/cm. Such a pixelated metasurface can be utilized to calculate the absorption fingerprint of surface-absorbed molecules by correlating the reflectance of each exposed metapixel with the molecular absorption at the resonance frequency of the metapixel. In the absence of the analyte, the reflectance peaks of the metapixels are about unity making the calculation of the molecular absorption fingerprint straightforward. It should be noted that proposed meta-atom can be readily scaled to cover a broader frequency range. To evaluate the effect of light-matter interaction on the reflectance spectrum, we suppose that the pixelated metasurface is coated with a thin layer of protein. After coating the meta-atoms with a 2.5 nm thick protein layer [26] with a permittivity model extracted from [49], the reflectance of the pixelated metasurface is calculated for the $y$-polarized incident field which is shown in Fig. 10(a). Here, the meta-atom is linearly scaled between $S$=1.00 and $S$=1.34 in 100 steps. For illustration purposes, only 100 steps are used. However, high quality factor of the deigned meta-atoms allows a higher number of scaling steps. The permittivity model of the protein used in calculations is displayed in Fig. 10(b). The absorption fingerprint of the protein is calculated by comparing the reflectance envelopes of the metasurface before and after exposure to the protein. The calculated absorption fingerprint is shown in Fig. 10(c). The absorption of the analyte calculated from the peak reflectance envelope has a good agreement with the absorption signature of the biomolecular analyte. It should be noted that the resonance frequency shift upon exposure to the protein is lower than 0.5 1/cm while the difference between the resonance frequency in the adjacent metpixels is about 3 1/cm for 100 scaling steps. Therefore, the resonances in the metapixel do not overlap. Nevertheless, this frequency shift has a negligible effect on the absorption fingerprint retrieval since we are relying on the reflectance envelope in our calculations. We should point out that the designed meta-atom is not suitable for the method relying on controlling the incidence angle of the source described in Introduction [36].

At oblique incidence, multiple resonances appear in the reflectance spectrum of the proposed meta-atom complicating the absorption fingerprint calculations.

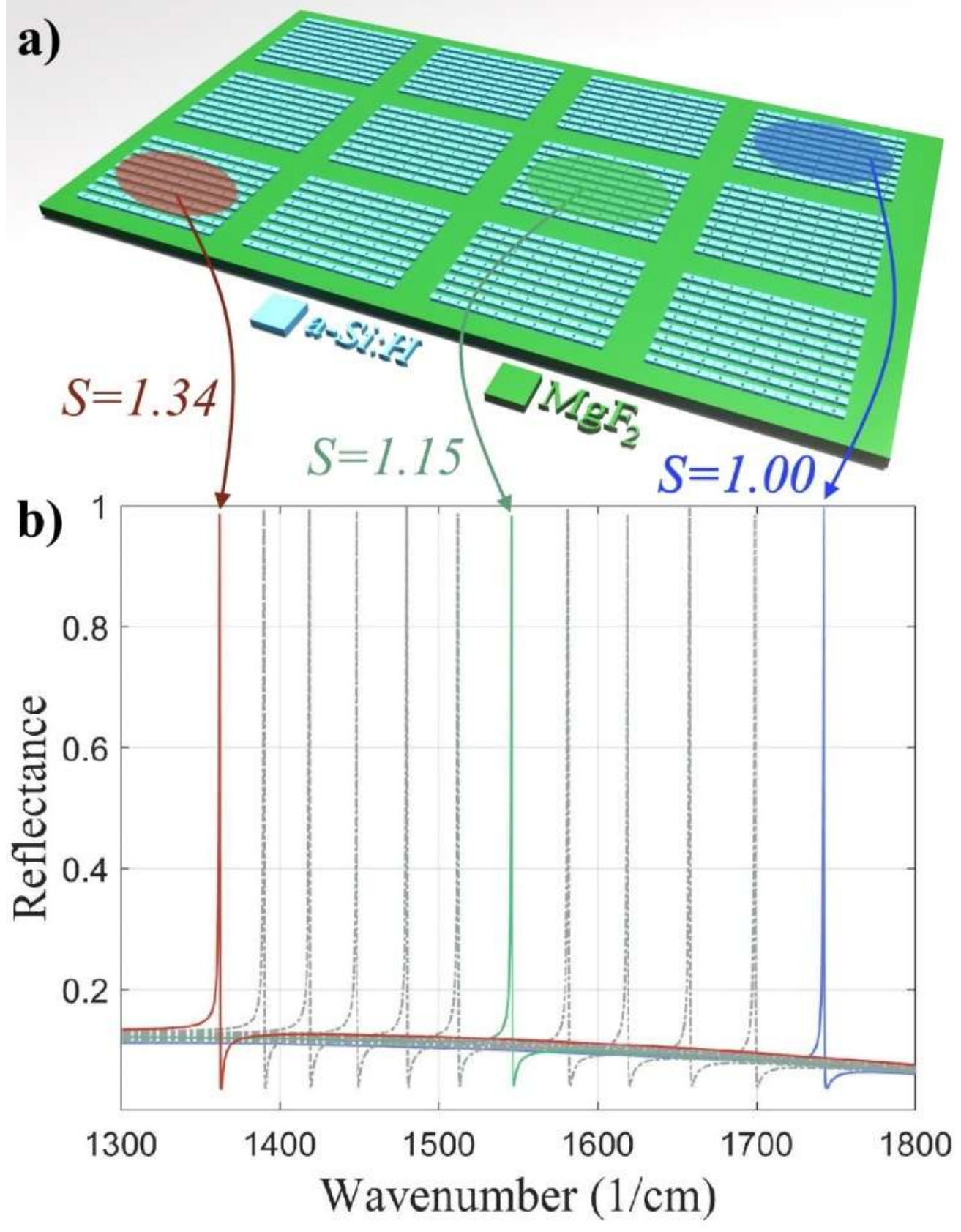

Fig. 9. a) Pixelated metasurface where each metapixel's resonance is tuned to a specific frequency. The unit cell and the meta-atoms are scaled by a factor of $S$ to provide an unique resonance in each metapixel. b) Simulated reflectance spectra of the metapixel with different scaling factors of $S$ for normally incident $y$-polarized wave and without the presence of the analyte.

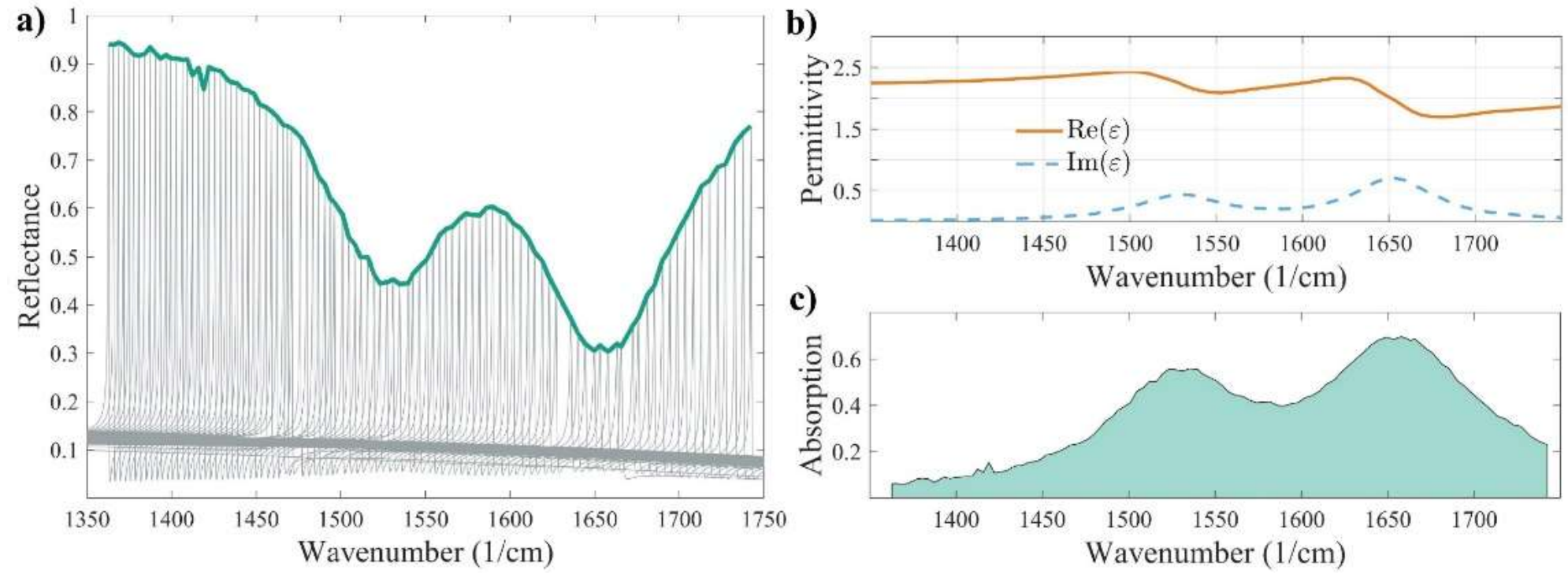

Fig. 10. a) Reflectance spectra of the pixelated metasurface after coating it with protein. b) The permittivity of the protein is redrawn from the data provided in [49]. c) Absorption fingerprint of the protein calculated from the reflectance envelopes before and after analyte coating.

## 4. Conclusion

In summary, the designed meta-atom with a quality factor larger than 1890 paves the way to design pixelated metasurfaces with smaller scaling steps without overlapping resonances. The performance of the pixelated metasurface to detect the absorption signature of surface-bound analyte is investigated. The numerical simulations show that the absorption signature of the target analyte molecules can be extracted by comparing the reflection spectra of the metasurface before and after exposing it to the analyte. In comparison to previous studies, the presented metasurfaces provide higher precision in retrieving molecular fingerprints.

**Disclosures.** The authors declare no conflicts of interest.

**Data availability.** Data underlying the results presented in this paper are not publicly available at this time but may be obtained from the authors upon reasonable request.